# Perfect Coulomb drag and exciton transport in an excitonic insulator

Ruishi Qi[1,2,†], Andrew Y. Joe[1,2,†,*], Zuocheng Zhang[1], Jingxu Xie[1,2], Qixin Feng[1,2], Zheyu Lu[1,2], Ziyu Wang[1,3], Takashi Taniguchi[4], Kenji Watanabe[5], Sefaattin Tongay[6], Feng Wang[1,2,7,*]

[1] Department of Physics, University of California, Berkeley, CA 94720, USA

[2] Materials Sciences Division, Lawrence Berkeley National Laboratory, Berkeley, CA 94720, USA

[3] School of Physics, Xi'an Jiaotong University, Xi'an 710049, China.

[4] Research Center for Materials Nanoarchitectonics, National Institute for Materials Science, 1-1 Namiki, Tsukuba 305-0044, Japan

[5] Research Center for Functional Materials, National Institute for Materials Science, 1-1 Namiki, Tsukuba 305-0044, Japan

[6] School for Engineering of Matter, Transport and Energy, Arizona State University, Tempe, AZ 85287, USA

[7] Kavli Energy NanoSciences Institute, University of California Berkeley and Lawrence Berkeley National Laboratory, Berkeley, CA 94720, USA.

† These authors contributed equally.

* To whom correspondence should be addressed: andrew.joe@ucr.edu, fengwang76@berkeley.edu

**Strongly coupled two-dimensional electron-hole bilayers can give rise to novel quantum Bosonic states: electrons and holes in electrically isolated layers can pair into interlayer excitons, which can form a Bose-Einstein condensate below a critical temperature at zero magnetic field. This state is predicted to feature perfect Coulomb drag, where a current in one layer must be accompanied by an equal but opposite current in the other, and counterflow superconductivity, where the excitons form a superfluid with zero viscosity. Electron-hole bilayers in the strong coupling limit with an excitonic insulator ground state have been recently achieved in semiconducting transition metal dichalcogenide heterostructures, but direct electrical transport measurements remain challenging. Here we use a novel optical spectroscopy to probe the electrical transport of correlated electron-hole fluids in $MoSe_2$/hBN/$WSe_2$ heterostructures. We observe perfect Coulomb drag in the excitonic insulator phase up to a temperature as high as ~15K. Strongly correlated electron and hole transport is also observed at unbalanced electron and hole densities, although the Coulomb drag is not perfect anymore. Meanwhile, the counterflow resistance of interlayer excitons remains finite. These results indicate the formation of an exciton gas in the excitonic insulator which does not condensate into a superfluid at low temperature. Our work also demonstrates that dynamic optical spectroscopy provides a powerful tool for probing novel exciton transport behavior and possible exciton superfluidity in correlated quantum electron-hole fluids.**

# Main

An electron-hole bilayer – a two-dimensional electron gas (2DEG) and a two-dimensional hole gas (2DHG) coupled together by Coulomb interactions while remaining electrically isolated – provides a highly tunable platform to study strongly correlated electron-hole fluids. In the strong coupling regime where the interlayer distance is small compared to intralayer particle spacing, the electrons and holes in adjacent layers pair into indirect excitons, and the system is expected to host novel quantum Bosonic states including correlated excitonic insulators[1,2], exciton supersolids[3], high-temperature bilayer exciton Bose-Einstein condensates (BEC)[2,4–6] and superfluids[7–10]. Such bilayer exciton condensates are characterized by a gapped energy spectrum, spontaneous interlayer phase coherence and dissipationless exciton transport[4,11]. The interlayer coherence is supposed to manifest itself as counterflow superconductivity with perfect Coulomb drag in a counterflow transport measurement – a current in one layer must be accompanied by an equal but opposite current in the other[12,13], and the longitudinal resistance vanishes.

Evidence of exciton condensation and perfect Coulomb drag has been reported in quantum Hall bilayers in semiconductor double quantum wells[12,14] and graphene systems[15,16], but the formation of quasi-electrons and quasi-holes in the quantized Landau levels requires a strong external magnetic field and the excitons only form at very low temperatures[17,18]. Recently, the research focus has shifted towards searching for an exciton condensate in electron-hole bilayers in the absence of a magnetic field[1,4,8,19].

Electron-hole bilayers in semiconducting transition metal dichalcogenide (TMD) heterostructures have attracted special interest due to the strong Coulomb interaction and large exciton binding energy (hundreds of meV)[20,21]. Consequently, the strong coupling regime becomes accessible[19,22]. This system provides a highly tunable platform to study intriguing transport behaviors of quantum exciton fluids. When the bandgap energy is electrically tuned below the exciton binding energy, strongly correlated excitonic insulator states have been experimentally achieved[1,23]. Theoretical study of such systems predicts an exciton BEC that persists until a very high temperature Berezinskii–Kosterlitz–Thouless (BKT) transition (a fraction of the exciton binding energy)[8,11]. Below this temperature, the interlayer excitons are expected to form an exciton superfluid, in which the exciton transports with zero viscosity, similar to the dissipationless transport of Cooper pairs in conventional superconductors. Electrical transport measurements are therefore highly desired to observe these exotic quantum states. However, most semiconducting TMDs have poor electrical contact to common metals due to a large Schottky barrier, which makes electrical transport measurements challenging. Therefore, experimental measurement of Coulomb drag and exciton transport remains elusive to date.

Here we develop a novel optical technique to quantitatively measure exciton transport behavior without the need to pass any current through contacts. This allows Coulomb drag measurements and accurate determination of exciton flow resistance. For the first time, we show that a TMD electron-hole bilayer features perfect Coulomb drag at equal electron and hole densities, which becomes non-perfect but remains very strong when additional charges are present. Surprisingly, our exciton transport measurements show the absence of an exciton superfluid for temperatures down to 2K, which is in contradiction to the theoretical predictions.

## Optical measurement of resistance and Coulomb drag

Figure 1a schematically shows the device structure and the experimental setup. We choose $MoSe_2$ as the electron layer and $WSe_2$ as the hole layer for their type-II band alignment. They are separated by a thin hBN tunnelling barrier to ensure equilibrium electron-hole fluids with ultralong exciton lifetime on the order of a second[23]. The heterostructure is encapsulated by dielectric hBN on both sides and gated by few-layer graphene top gate (TG) and two back gates (BG1 and BG2). This dual-gated electron-hole bilayer device has two regions controllable by separate BGs, each easily tunable with electrical voltages. We begin by considering region 1. We keep the electron layer grounded ($V_e = 0$) and apply voltages on the gates and the hole layer. The gate voltage $V_G \equiv V_{TG} + V_{BG1}$ tunes the Fermi-level and thus the net charge density (electron-hole imbalance). The hole-electron voltage difference $V_h - V_e$ and the vertical electric field $V_{TG} - V_{BG1}$ both tune the band alignment. Therefore, the effective bias voltage $V_B \equiv (V_h - V_e) + \frac{t_m}{t_t + t_m + t_b}(V_{TG} - V_{BG1})$, where $t_t, t_m, t_b$ are top, middle and bottom hBN thicknesses in region 1 respectively, controls the type-II band gap. Fig. 1b is an optical image of such a device (D1), whose detailed structure is given in Extended Data Fig. 1. While we define all the electrical voltages using region 1, we scale the d.c. voltage on BG2 appropriately such that region 1 and region 2 remain in the same gating and electric field conditions, and thus the electron and hole densities remain homogeneous (details in Methods). Unless otherwise specified, the data shown below is taken from D1 at a temperature $T = 2K$. Similar data can be reproduced in device D2, as summarized in Extended Data Fig. 2.

Using reflectance spectroscopy, we are able to determine the electron and hole densities as a function of gate and bias voltages[23]. Figure 1c shows the charge doping phase diagram as a function of $V_G$ and $V_B$, where red and green channels of the false color map encode the density of electrons ($n_e$) and holes ($n_h$) respectively. At low bias voltages, the system has a finite type-II band gap, so only one type of charge can enter the system at a time. When the bias voltage exceeds the gap energy of approximately 1.51eV, electrons and holes enter the system at the same time, forming correlated electron-hole fluids that consists of interlayer excitons and possibly extra charges[1,23]. In particular, when the bias voltage reduces the single-particle band gap to be smaller than the exciton binding energy but still nonzero, an excitonic insulator phase emerges that consists of only interlayer excitons, while no extra unpaired charges are allowed to exist due to the finite single-particle gap[1,23]. The charge compressibility map[23] in Fig. 1d reveals this charge-incompressible state at finite and equal electron and hole densities (triangle region). In this phase, perfect interlayer correlation is expected. The Coulomb drag is therefore expected to become perfect – any current flow in one layer necessarily needs to be accompanied by an equal but opposite current in the other layer.

In standard electrical transport measurements, multiple contacts are made to each layer to pass currents and measure the resulting voltage drop. However, the contact resistance for TMD layers are usually orders of magnitude larger than the sheet resistance, and therefore such transport measurements are very challenging. To overcome this problem, we do not drive currents through contacts. Instead, we capacitively drive current between the two heterostructure regions[24] in our device by applying a small voltage modulation $U = 5mV_{rms}$ at angular frequency $\omega$ between BG1 and BG2, as illustrated in Fig. 1a. This a.c. voltage generates an oscillating potential that drives charges/excitons to flow back and forth, leading to an a.c. particle density change at $\omega$. We then optically detect the density change. The optical absorption of the TMD layers is known to depend sensitively on its local charge density[25,26]. For example, Figure 1e shows the density dependence of the device reflectivity spectrum near the MoSe$_2$ A exciton wavelength. With increasing electron density, the intralayer exciton peak loses its oscillator strength, and an additional absorption peak, commonly known as trions, appears at lower energy[25–27]. We focus a monochromatic laser probe at region 2 and use an avalanche photodiode (APD) to read out the reflected light intensity, whose

a.c. component is proportional to the local electron density oscillation $\Delta n_e$. Similarly, when the laser wavelength is tuned to the WSe$_2$ absorption peak, $\Delta n_h$ can be measured. Since the density change is directly proportional to the current flow, our technique optically probes the electrical transport behavior.

Figure 1f shows the effective circuit of our measurement. When the gate modulation is slow, the modulation of electron and hole density is determined by the geometric and quantum capacitance in the coupled system. With increasing modulation frequency $\omega$, the charges/excitons need to move faster to respond to the gate modulation, until a characteristic cutoff frequency beyond which the motion of the particles will be limited by their mobility and become ineffective. This cutoff frequency is determined by the $RC$ constant of the circuit, yielding the longitudinal resistance $R$.

## Perfect Coulomb drag

Figure 2a-b shows the real part of the hole ($\Delta n_h$) and electron ($\Delta n_e$) density oscillation averaged over the low-frequency voltage modulation regime ($\omega = 0.82 - 41$ kHz). The imaginary part is always essentially zero at low frequency (see Extended Data Fig. 3 for a complete data set). At low bias voltages ($V_B$<1.51V), the system forms a 2DEG or a 2DHG with only one type of charge present. The gate modulation will directly drive the charges in the active layer. The holes (electrons) have a well-defined positive (negative) response, due to their opposite charge. When a high bias voltage closes the gap and both electrons and holes are present, the experiment turns into a Coulomb drag measurement, where the holes are directly capacitively coupled to the back gate modulation and drag the electrons with them. If there were no interlayer coupling and both layers acted as perfect metals, the hole layer would completely screen the gate modulation and the electron layer would have no response. However, the experimental data shows that $\Delta n_e$ remains very strong when both layers are doped, indicating strong interlayer interactions in the electron-hole bilayer. Such strong Coulomb drag, where the drag signal is comparable to the drive signal, is unusual, as most coupled bilayer systems have Coulomb drag that is on the order of a percent[28,29]. We note that the sign of $\Delta n_e$ becomes positive in this regime, in contrast to the negative response when there are only electrons in the system. This supports the picture of electrons moving together with holes. For the hole response, the sign remains the same regardless of the electron doping, but the magnitude changes for different doping conditions.

Figure 2c shows the drag ratio, $\eta = \Delta n_e/\Delta n_h$. Noticeably, the drag ratio features a very strong enhancement when the electron and hole densities are equal. In the excitonic insulator phase, we observe significantly increased drag response and decreased drive response. The drag ratio approaches 1 in this triangle region, which is consistent with the physical picture of an excitonic insulator phase that does not allow the existence of any unpaired charge. Thus, the motion of a hole must be accompanied by an electron.

We now focus on a horizontal linecut at constant bias $V_B = 1.52$V (Fig. 3a). In this linecut, constant bias voltage leads to fixed $n_e + n_h \approx 0.3 \times 10^{12}$cm$^{-2}$ and the gate voltage tunes the electron-hole imbalance $n_e - n_h$. The drag signal peaks at $n_e = n_h$ and decreases rapidly when additional charges are present. Meanwhile, the drive layer has reduced response in the excitonic insulator region, outside which the drive signal varies smoothly.

The frequency-dependent density change of the hole and the electron layers are displayed in Fig. 3b-c. The responses from both layers remain constant for low frequencies below 100 kHz, and start to decay after the modulation frequency exceeds the characteristic $RC$ time constant. The gate dependence of the signal shows an abrupt change across the net charge neutrality point, indicating strong excitonic effects in the electron-hole bilayer. Figs. 3d-f displays frequency sweeps at three typical doping conditions, with $n_e < n_h$, $n_e = n_h$, and $n_e > n_h$ respectively. We observe qualitatively distinct behavior among them. When $n_e = n_h$, the electron and hole responses become identical, with the same amplitude, same phase (coded by the color of the scatter points), and same cutoff frequency. This demonstrates perfect Coulomb drag in dynamic transport, where the drive current and drag current are identical at all frequencies. When either type of additional charge is present, the active and passive layer responses are no longer identical.

## Charge and exciton transport

To quantitatively understand the transport behavior of the correlated electron-hole fluids, we use the effective circuit shown in Fig. 1f to model the charge and exciton transport. Let $\phi_{ij}$ and $\mu_{ij}$ ($i \in \{e, h\}, j = 1,2$) denote the electric potential and chemical potential of layer $i$ and region $j$. In the case of strong interlayer coupling, the current in the electron and hole layers $(I_e, I_h)$ and the electrochemical potential drops ($\Delta\phi_i \equiv \phi_{i1} - \phi_{i2}$, $\Delta\mu_i \equiv \mu_{i1} - \mu_{i2}$) are related by a $2 \times 2$ conductance matrix $G$

$$\begin{bmatrix} I_e \\ I_h \end{bmatrix} = \begin{bmatrix} G_{11} & G_{12} \\ G_{21} & G_{22} \end{bmatrix} \cdot \begin{bmatrix} \Delta\phi_e + \Delta\mu_e \\ \Delta\phi_h + \Delta\mu_h \end{bmatrix} \quad (1)$$

The conductance matrix can potentially take contributions from unpaired electrons, unpaired holes, and interlayer excitons. Ignoring the weak frictional drag between unpaired charges, unpaired electrons in the top layer give a contribution $\begin{bmatrix} 1/R_e & 0 \\ 0 & 0 \end{bmatrix}$ to the conductance matrix, and similarly unpaired holes will give $\begin{bmatrix} 0 & 0 \\ 0 & 1/R_h \end{bmatrix}$, where $R_e$ and $R_h$ are the effective longitudinal resistance for the electrons and the holes. These two terms both take a diagonal form and do not contribute to the drag effect. The interlayer exciton transport, however, will give an additional term that takes the form $\frac{1}{R_x}\begin{bmatrix} 1 & -1 \\ -1 & 1 \end{bmatrix}$. This is because interlayer exciton motion is driven by the difference in the potential drops $\Delta\phi_h + \Delta\mu_h - \Delta\phi_e - \Delta\mu_e$, and will lead to an opposite current in the two layers.

The electric potentials $\phi_{ij}$ and currents $\{I_e, I_h\}$ are related to the gate voltage modulation by basic circuit laws in this capacitor system. The chemical potential drops are related to the charge density changes by the quantum capacitance matrix $C_Q = \begin{bmatrix} \frac{\partial \mu_e}{\partial n_e} & \frac{\partial \mu_e}{\partial n_h} \\ \frac{\partial \mu_h}{\partial n_e} & \frac{\partial \mu_h}{\partial n_h} \end{bmatrix}$ (details in Methods). With these equations together, the effective circuit model is fitted to the experimental $\Delta n_e, \Delta n_h$ to extract the three resistances $R_e$, $R_h$ and $R_x$. The fitting agrees well with the experiment for both the amplitude and the phase of $\Delta n_e$ and $\Delta n_h$ (solid lines in Fig. 3d-f).

Fig. 3g plots the fitted resistances as a function of $V_G$. The exciton resistance is on the order of hundreds of kiloohms and only has a weak gate dependence across the charge neutrality. The unpaired electron and hole resistance change dramatically across net charge neutrality, and they do not become conductive at the same time. When $n_e < n_h$, $R_e$ is very large and beyond our measurement range, but when $n_e$ exceeds $n_h$ it quickly drops below 1MΩ and decreases with increasing electron density. Similar behavior is observed for the hole resistance. This observation indicates that in the low-density regime the minority carrier cannot exist in the unpaired form, maximizing the number of interlayer excitons. At net charge neutrality, all the electrons and holes pair into bound states of indirect excitons, and the motion of both free charges becomes frozen, leading to perfect Coulomb drag where the current flow can only come from the exciton motion. The absence of unpaired electron/hole contribution to the conductance matrix results in the divergence of drag resistance, as the exciton conductance matrix is not invertible.

Next, we examine the exciton phase diagram at net charge neutrality $n_e = n_h = n$. Fig. 4a-b show the hole and electron density change as a function of pair density $n$. At low densities, we observe the same response from the two layers, and the drag ratio (Fig. 4c, right axis) is close to 1 for $n \lesssim 0.3 \times 10^{12} \text{cm}^{-2}$. With increased density, the drag ratio gradually decreases due to the reduced single-particle gap. It remains considerable even after the Mott density $n_M \approx 0.8 \times 10^{12} \text{cm}^{-2}$.[1,23] The drag ratio shown in Fig. 4d also decreases with elevated temperature. At low densities, the Coulomb drag remains almost perfect (>85%) up to ~15K, which is orders of magnitude higher than semiconductor quantum wells and graphene-based systems at high magnetic field[12,15]. The drag ratio decreases slowly with higher temperature, indicating the coexistence of excitons, unpaired electrons, and unpaired holes at finite temperatures due to finite thermal energy. After complete thermal melting of interlayer excitons at ~70K,[23] the drag signal at 80K reduces significantly to only ~20%. The drag signal does not completely disappear after the quantum dissociation at high exciton density or thermal melting at high temperature, suggesting a considerable drag effect in the electron-hole plasma phase due to strong Coulomb attractions between electrons and holes.

## Discussion and outlook

Fitting the electron and hole responses to the effective circuit model with exciton transport term yields the exciton resistance $R_x$ as a function of density $n$, as shown in Fig. 4e. With increasing density, the exciton resistance decreases slightly faster than $1/n$ scaling (dotted lines), suggesting only a small increase in the exciton mobility at higher doping. A superfluid transition is not observed, which can also be seen from the high-frequency decay in Figs. 4a-b. We performed measurements with different driving voltages and hence different exciton current flow (Extended Data Fig. 4) and conclude that the exciton transport behavior is linear. Extended Data Fig. 5 compares the resistance of the exciton gas, the 2DHG in the WSe$_2$ layer, and the 2DEG in the MoSe$_2$ layer as a function of carrier density. At similar carrier density, the excitons here are always several times more resistive than the 2DHG in the WSe$_2$ layer. The resistance of excitons is overall closer to the resistance of the 2DEG in the MoSe$_2$ layer, but they show a different density dependence. The resistance of the 2DEG increases significantly faster than $1/n_e$ at low density, potentially due to Wigner crystallization[30] or unscreened charge defects. At higher density the excitons become more resistive than both unpaired electrons and unpaired holes. The different scaling behavior suggests a different scattering mechanism for interlayer excitons compared with unpaired charges.

Temperature dependence of the exciton resistance (Fig. 4f) reveals that the resistance decreases with increasing temperature. Contrary to previous theoretical predictions of superfluidity over a broad temperature range in this system[6,8,11], we do not observe any signature of an exciton superfluid down to our base temperature $T = 2K$. This could possibly be due to disorders in the samples that might destroy the phase coherence between the excitons.

In conclusion, we have demonstrated a powerful technique that can optically measure the Coulomb drag effect. We observe perfect Coulomb drag in the excitonic insulator phase of the electron-hole bilayer system without any external magnetic field. The transport behavior of a stable interlayer exciton fluid is measured for the first time. Our results establish TMD-based electron-hole bilayers as a promising platform for novel exciton-based electronic devices. Despite negative results in the search for exciton superfluidity in TMD bilayers, our work paves the way for further study of counterflow superconductivity and exciton condensates when the sample quality improves further.

## Methods

**Device fabrication.** We use a dry-transfer method based on polyethylene terephthalate glycol (PETG) stamps to fabricate the heterostructures. Monolayer MoSe$_2$, monolayer WSe$_2$, few-layer graphene and hBN flakes are mechanically exfoliated from bulk crystals onto Si substrates with a 90-nm-thick SiO$_2$ layer. We use ~20 nm hBN as the gate dielectric and 2-3 nm thin hBN as the interlayer spacer. A 0.5 mm thick clear PETG stamp is employed to sequentially pick up the flakes at 65-75 °C. The whole stack is then released onto a high resistivity Si substrate with a 90 nm SiO$_2$ layer at 95-100 °C, followed by dissolving the PETG in chloroform at room temperature for one day. Electrodes (50 nm Au with 5 nm Cr adhesion layer) are defined using photolithography (Durham Magneto Optics, MicroWriter) and electron beam evaporation (Angstrom Engineering).

Extended Data Fig. 1 shows detailed device structures. We use two graphite back gates that are vertically stacked with a thin hBN dielectric in between to ensure homogeneous d.c. doping without any ungated region between two back gates. All the voltages are first defined in region 1. The d.c. voltage on BG2 is dependent on BG1 such that the densities in the two regions remain uniform. The scaling for device D1 is $V_{BG2} - V_h = 1.44(V_{BG1} - V_h) + 0.65V$, where the slope and the intercept are determined experimentally that achieves the best matching of doping boundaries for both the electron and the hole layers. Extended Data Figs. 1e-f give a comparison of the doping phase diagram in the two regions after this scaling, which matches very well down to mV level.

To ensure equilibrium electron-hole fluids, we improve the contacts to TMD layers with specially designed contact regions. For device D1, we follow ref. [1] and insert a thick hBN layer in the contact region (region 0, see Extended Data Fig. 1a). With larger interlayer spacing in this region, the vertical electric field $V_{TG} - V_{BG}$ will easily close the type-II band gap and make the contact region heavily doped. For device D2, we misalign the top and back gates with the heterostructure such that the MoSe$_2$ contact region is only covered by the top gate while the WSe$_2$ contact region is only gated by the back gate (Extended Data Fig. 1c). This also makes the contact region heavily doped. In both designs, the contact resistance due to large Schottky barriers are reduced to eliminate any observable doping hysteresis in the voltage ranges of interest. However, the contact resistance is still large so that for the frequency range relevant in this paper ($\omega = 10^2\text{-}10^9$Hz), the electron and hole contacts are both frozen and the TMD layers are effectively floating. The charge and exciton transport is therefore only between heterostructure region 1 and region 2, not from the electrode to the heterostructure.

**Optical measurements.** The optical measurements are performed in an optical cryostat (Quantum Design, OptiCool) with a temperature down to 2 K (nominal). The reflection spectroscopy is performed with a supercontinuum laser (Fianium Femtopower 1060 Supercontinuum Laser) as the light source.

In optical transport measurements, the gate voltage modulation is applied by an arbitrary waveform generator (Siglent SDG6022X). A small voltage modulation with $U_{\text{rms}} = 5$mV is used to ensure minimal perturbation to the system, unless otherwise specified. Keithley 2400 or 2450 source meters are used for applying other gate and bias voltages and monitoring the leakage current.

The monochromatic laser probe is either the supercontinuum laser with the wavelength selected by a reflective grating and an iris, or a diode laser with its wavelength finely tuned by a thermoelectric cooler. The laser is focused on the sample by a 20× Mitutoyo objective with ~1.5µm beam size. We choose a low incident laser power (<20 nW) to minimize photodoping effects. The reflected light is collected by an APD (Thorlabs APD 410A) and analyzed using a lock-in amplifier (Stanford Research SR865A). For angular frequency below 25MHz, the APD voltage output is directly analyzed with the lock-in amplifier that is locked to the function generator output frequency. For higher frequencies ($\omega$>25MHz), we use another scheme to convert the high-frequency reflectivity oscillation to a low-frequency signal that can be easily collected by the APD and analyzed by the lock-in amplifier. The voltage input of the laser diode is modulated at frequency $\omega' = \omega + \delta\omega$. This leads to an incident laser intensity oscillating at a frequency slightly higher than the gate voltage modulation frequency. The multiplication of incident laser power oscillating at $\omega + \delta\omega$ and the device reflectivity oscillating at $\omega$ generates a low frequency component at $\delta\omega$ for the reflected light intensity. The lock-in amplifier is then locked to the frequency difference $\delta\omega = 6.5$kHz.

The lock-in amplifier output gives the a.c. part of the APD voltage. The APD d.c. output voltage is recorded by a data acquisition card (NI USB-6212) analogue input. The ratio between the lock-in output and the APD d.c. voltage gives the relative reflectivity change $\Delta r/r$, where $r$ denotes the reflectivity at the laser wavelength and is a function of the charge density $n$. The density change $\Delta n$ can be then determined by

$$\frac{\Delta r}{r} = \Delta n \times \frac{\mathrm{d}r(n)}{r\mathrm{d}n}$$

The sensitivity $\frac{\mathrm{d}r(n)}{r\mathrm{d}n}$ is directly derived from the gate dependence of the APD d.c. output.

**Effective circuit model**. The conductance matrix $G$ is related to the resistance of the free electrons, free holes, and excitons by

$$G = \frac{1}{R_e}\begin{bmatrix} 1 & 0 \\ 0 & 0 \end{bmatrix} + \frac{1}{R_h}\begin{bmatrix} 0 & 0 \\ 0 & 1 \end{bmatrix} + \frac{1}{R_x}\begin{bmatrix} 1 & -1 \\ -1 & 1 \end{bmatrix}$$

The current $\begin{bmatrix} I_e \\ I_h \end{bmatrix}$ is driven by the electrochemical potential difference in the two regions $\begin{bmatrix} \Delta\phi_e + \Delta\mu_e \\ \Delta\phi_h + \Delta\mu_h \end{bmatrix}$ (Equation (1) in the main text).

First, the electric potentials $\phi$ and the applied modulation voltage are related by the Kirchhoff circuit laws

$$i\omega(C_{t1} + C_{m1})\phi_{e1} - i\omega C_{m1}\phi_{h1} + I_e = 0$$

$$i\omega(C_{t2} + C_{m2})\phi_{e2} - i\omega C_{m2}\phi_{h2} - I_e = 0$$

$$i\omega(C_{b1} + C_{m1})\phi_{h1} - i\omega C_{m1}\phi_{e1} + I_h = 0$$

$$i\omega(C_{b2} + C_{m2})\phi_{h2} - i\omega C_{m2}\phi_{e2} - I_h = i\omega C_{m2} U$$

Here $C_{ij}$ is the geometric capacitance at the top, middle, and bottom of each heterostructure region determined by the heterostructure area $A_1, A_2$ and hBN thicknesses $t_t, t_m, t_b$. Subscript $i \in \{t, m, b\}$ denotes top, middle, or bottom and $j \in \{1,2\}$ denotes region 1 or 2. The dielectric constant of hBN is 4.2.

Second, the density changes in region 1 and region 2 are $\begin{bmatrix} I_e/i\omega e A_1 \\ -I_h/i\omega e A_1 \end{bmatrix}$ and $\begin{bmatrix} -I_e/i\omega e A_2 \\ I_h/i\omega e A_2 \end{bmatrix}$ respectively. Here $e$ is the elementary charge. This leads to a chemical potential difference between them as

$$\begin{bmatrix} \Delta\mu_e \\ \Delta\mu_h \end{bmatrix} = \begin{bmatrix} \frac{\partial\mu_e}{\partial n_e} & \frac{\partial\mu_e}{\partial n_h} \\ \frac{\partial\mu_h}{\partial n_e} & \frac{\partial\mu_h}{\partial n_h} \end{bmatrix} \begin{bmatrix} I_e/i\omega e(A_1 + A_2) \\ -I_h/i\omega e(A_1 + A_2) \end{bmatrix}$$

For a given set of conductance matrix and quantum capacitance matrix, the equations above determine $\phi_{ij}$, which then gives the charge density changes $(\Delta n_e, \Delta n_h)$ in the electron and hole layers in region 2 by

$$-eA_2\Delta n_e = C_{t2}\phi_{e2} + C_{m2}(\phi_{e2} - \phi_{h2})$$

$$eA_2\Delta n_h = C_{b2}(\phi_{h2} - U) + C_{m2}(\phi_{h2} - \phi_{e2})$$

In summary, the effective circuit model above can determine $(\Delta n_e, \Delta n_h)$ from a set of conductance matrix and quantum capacitance matrix. By fitting the experimental $(\Delta n_e, \Delta n_h)$ to the model, we can extract the electron, hole, and exciton resistances $R_e, R_h, R_x$.

**References**
1. Ma, L. *et al.* Strongly correlated excitonic insulator in atomic double layers. *Nature* **598**, 585–589 (2021).
2. Zeng, Y. & MacDonald, A. H. Electrically controlled two-dimensional electron-hole fluids. *Phys. Rev. B* **102**, 085154 (2020).
3. Conti, S. *et al.* Chester Supersolid of Spatially Indirect Excitons in Double-Layer Semiconductor Heterostructures. *Phys. Rev. Lett.* **130**, 057001 (2023).

4. Wu, F.-C., Xue, F. & MacDonald, A. H. Theory of two-dimensional spatially indirect equilibrium exciton condensates. *Phys. Rev. B* **92**, 165121 (2015).
5. Zhu, X., Littlewood, P. B., Hybertsen, M. S. & Rice, T. M. Exciton Condensate in Semiconductor Quantum Well Structures. *Phys. Rev. Lett.* **74**, 1633–1636 (1995).
6. Lozovik, Y. E., Kurbakov, I. L., Astrakharchik, G. E., Boronat, J. & Willander, M. Strong correlation effects in 2D Bose–Einstein condensed dipolar excitons. *Solid State Commun.* **144**, 399–404 (2007).
7. Pieri, P., Neilson, D. & Strinati, G. C. Effects of density imbalance on the BCS-BEC crossover in semiconductor electron-hole bilayers. *Phys. Rev. B* **75**, 113301 (2007).
8. Fogler, M. M., Butov, L. V. & Novoselov, K. S. High-temperature superfluidity with indirect excitons in van der Waals heterostructures. *Nat. Commun.* **5**, 4555 (2014).
9. Su, J.-J. & MacDonald, A. H. How to make a bilayer exciton condensate flow. *Nat. Phys.* **4**, 799–802 (2008).
10. Gupta, S., Kutana, A. & Yakobson, B. I. Heterobilayers of 2D materials as a platform for excitonic superfluidity. *Nat. Commun.* **11**, 2989 (2020).
11. Debnath, B., Barlas, Y., Wickramaratne, D., Neupane, M. R. & Lake, R. K. Exciton condensate in bilayer transition metal dichalcogenides: Strong coupling regime. *Phys. Rev. B* **96**, 174504 (2017).
12. Nandi, D., Finck, A. D. K., Eisenstein, J. P., Pfeiffer, L. N. & West, K. W. Exciton condensation and perfect Coulomb drag. *Nature* **488**, 481–484 (2012).
13. Narozhny, B. N. & Levchenko, A. Coulomb drag. *Rev. Mod. Phys.* **88**, 025003 (2016).
14. Butov, L. V., Zrenner, A., Abstreiter, G., Böhm, G. & Weimann, G. Condensation of Indirect Excitons in Coupled AlAs/GaAs Quantum Wells. *Phys. Rev. Lett.* **73**, 304–307 (1994).
15. Liu, X., Watanabe, K., Taniguchi, T., Halperin, B. I. & Kim, P. Quantum Hall drag of exciton condensate in graphene. *Nat. Phys.* **13**, 746–750 (2017).
16. Li, J. I. A., Taniguchi, T., Watanabe, K., Hone, J. & Dean, C. R. Excitonic superfluid phase in double bilayer graphene. *Nat. Phys.* **13**, 751–755 (2017).
17. Eisenstein, J. P. Exciton Condensation in Bilayer Quantum Hall Systems. *Annu. Rev. Condens. Matter Phys.* **5**, 159–181 (2014).
18. Eisenstein, J. P. & MacDonald, A. H. Bose–Einstein condensation of excitons in bilayer electron systems. *Nature* **432**, 691–694 (2004).
19. Wang, Z. *et al.* Evidence of high-temperature exciton condensation in two-dimensional atomic double layers. *Nature* **574**, 76–80 (2019).
20. Mak, K. F. & Shan, J. Photonics and optoelectronics of 2D semiconductor transition metal dichalcogenides. *Nat. Photonics* **10**, 216–226 (2016).
21. Wang, G. *et al.* Excitons in atomically thin transition metal dichalcogenides. *Rev. Mod. Phys.* **90**, 021001 (2018).
22. Jauregui, L. A. *et al.* Electrical control of interlayer exciton dynamics in atomically thin heterostructures. *Science* **366**, 870–875 (2019).
23. Qi, R. *et al.* Thermodynamic behavior of correlated electron-hole fluids in van der Waals heterostructures. Preprint at https://doi.org/10.48550/arXiv.2306.13265 (2023).
24. Regan, E. C. *et al.* Mott and generalized Wigner crystal states in WSe2/WS2 moiré superlattices. *Nature* **579**, 359–363 (2020).
25. Mak, K. F. *et al.* Tightly bound trions in monolayer MoS2. *Nat. Mater.* **12**, 207–211 (2012).

26. Ross, J. S. *et al.* Electrical control of neutral and charged excitons in a monolayer semiconductor. *Nat. Commun.* **4**, 1474 (2013).
27. Scuri, G. *et al.* Large Excitonic Reflectivity of Monolayer MoSe2 Encapsulated in Hexagonal Boron Nitride. *Phys. Rev. Lett.* **120**, 037402 (2018).
28. Kim, S. *et al.* Coulomb drag of massless fermions in graphene. *Phys. Rev. B* **83**, 161401 (2011).
29. Gramila, T. J., Eisenstein, J. P., MacDonald, A. H., Pfeiffer, L. N. & West, K. W. Mutual friction between parallel two-dimensional electron systems. *Phys. Rev. Lett.* **66**, 1216–1219 (1991).
30. Smoleński, T. *et al.* Signatures of Wigner crystal of electrons in a monolayer semiconductor. *Nature* **595**, 53–57 (2021).

## Data availability
The data that support the findings of this study are available from the corresponding author upon reasonable request.

## Acknowledgements

The optical spectroscopy of exciton transport measurements was supported by the AFOSR award FA9550-23-1-0246. The van der Waals heterostructure fabrication was supported by the U.S. Department of Energy, Office of Science, Office of Basic Energy Sciences, Materials Sciences and Engineering Division under contract no. DE-AC02-05-CH11231 (van der Waals heterostructures programme, KCWF16). S.T. acknowledges support from DOE-SC0020653, NSF CMMI 1933214, NSF mid-scale 1935994, NSF 1904716, NSF DMR 1552220 and DMR 1955889. K.W. and T.T. acknowledge support from the JSPS KAKENHI (Grant Numbers 21H05233 and 23H02052) and World Premier International Research Center Initiative (WPI), MEXT, Japan.

## Author contributions
F.W. conceived the research. R.Q. fabricated the devices with help from A.Y.J., J.X., Q.F., Z.W. and Z.L.. R.Q. and A.Y.J. performed the optical measurements assisted by Z.Z. and J.X.. R.Q., A.Y.J and F.W. analyzed the data. S.T. grew WSe$_2$ and MoSe$_2$ crystals. K.W. and T.T. grew hBN crystals. All authors discussed the results and wrote the manuscript.

## Competing interests
The authors declare no competing interests.

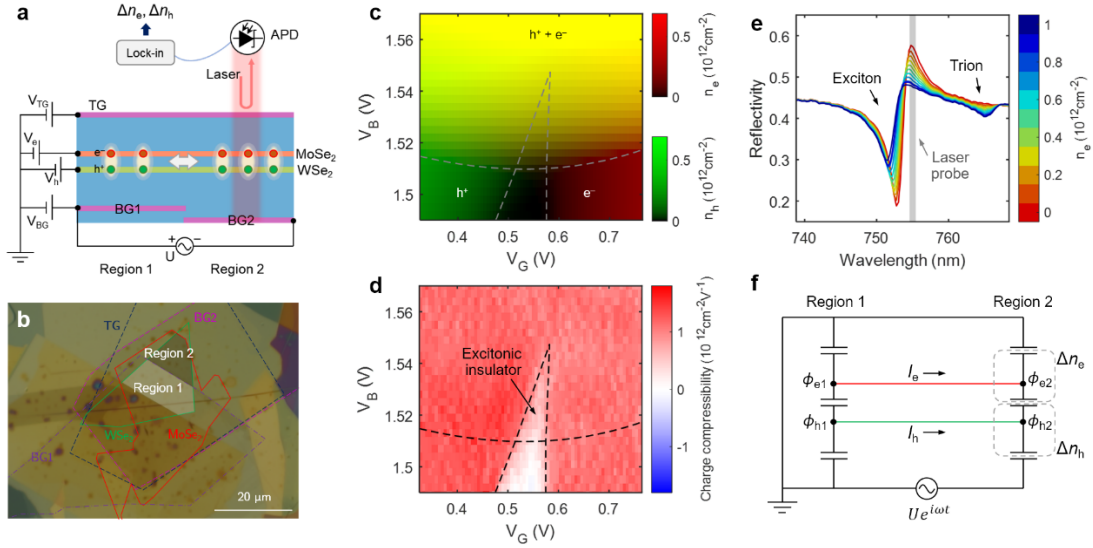

**Fig. 1 | Correlated electron-hole bilayer device.**
**a**, Schematic cross-section of D1, the MoSe$_2$/hBN/WSe$_2$ heterostructure device.
**b**, An optical image of device D1, with flake boundaries outlined.
**c**, Carrier doping phase diagram of the electron-hole bilayer in region 2. The red and green channel of the image show the electron density and the hole density respectively.
**d**, Charge compressibility from numerical partial derivative of electron-hole density imbalance with respect to gate voltage, $\partial(n_e - n_h)/\partial V_G$.
**e**, Device reflectivity spectrum as a function of electron doping density in the MoSe$_2$ layer. The gray line marks the wavelength of the laser probe.
**f**, Effective circuit model for transport measurements where $I_{e/h}$ is the current in the electron (hole) layers, $\phi_{ei}$ ($\phi_{hi}$) is the electric potential for region $i = 1,2$ for the electron (hole) layer, and $\Delta n_e$ ($\Delta n_h$) is the a.c. electron (hole) density change.

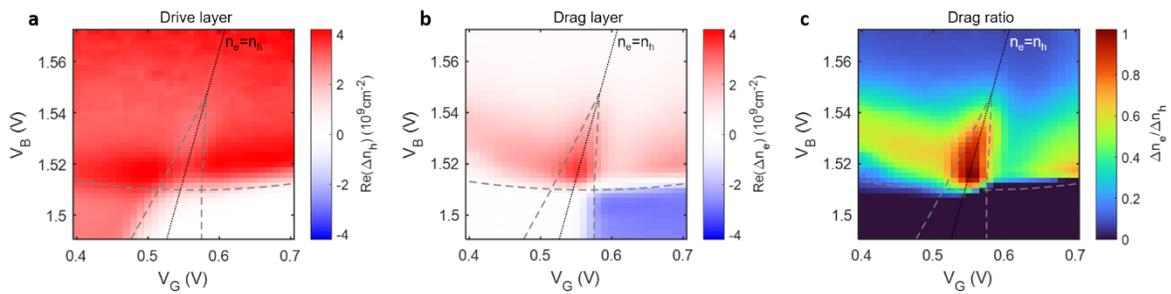

**Fig. 2 | Strong Coulomb drag in the electron-hole bilayer.**
**a-b**, Density change of the drive layer (a) and the drag layer (b) under a low frequency voltage modulation (averaged for $\omega = 82\mathrm{Hz} - 41\mathrm{kHz}$).
**c**, Drag ratio $\eta = \Delta n_e/\Delta n_h$ as a function of gate and bias voltages.

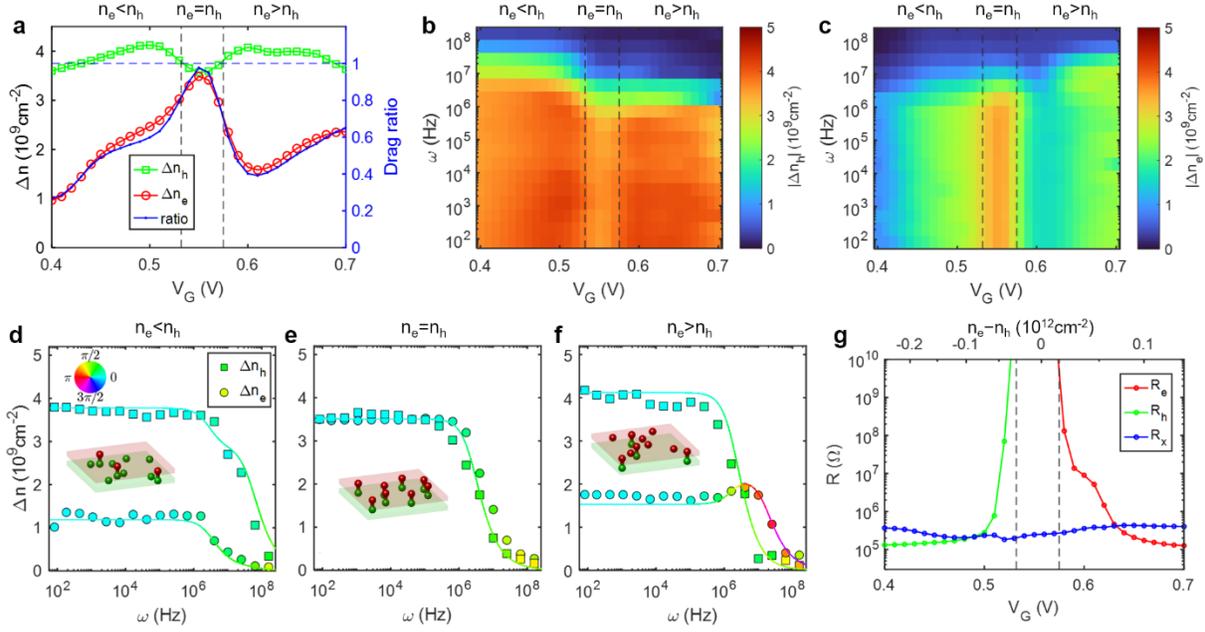

**Fig. 3 | Exciton and charge transport.**
**a**, Drive (hole) and drag (electron) layer density change as a function of gate voltage at constant bias $V_B = 1.52V$. Right axis: the ratio between $\Delta n_e$ and $\Delta n_h$, revealing perfect Coulomb drag at net charge neutrality.
**b-c**, Drive (hole) and drag (electron) layer density change under different driving frequencies $\omega$.
**d-f**, Frequency sweeps at $V_G = 0.42V, 0.55V$ and $0.63V$, corresponding to three typical doping conditions $n_e < n_h, n_e = n_h$ and $n_e > n_h$ respectively. Squares and circles are experimental data for $\Delta n_h$ and $\Delta n_e$. Solid lines, fitting results from the effective circuit model. The color of the markers and lines displays the phase of the density change.
**g**, Fitted resistance of free electron ($R_e$), free hole ($R_h$) and interlayer exciton ($R_x$), as a function of $V_G$.

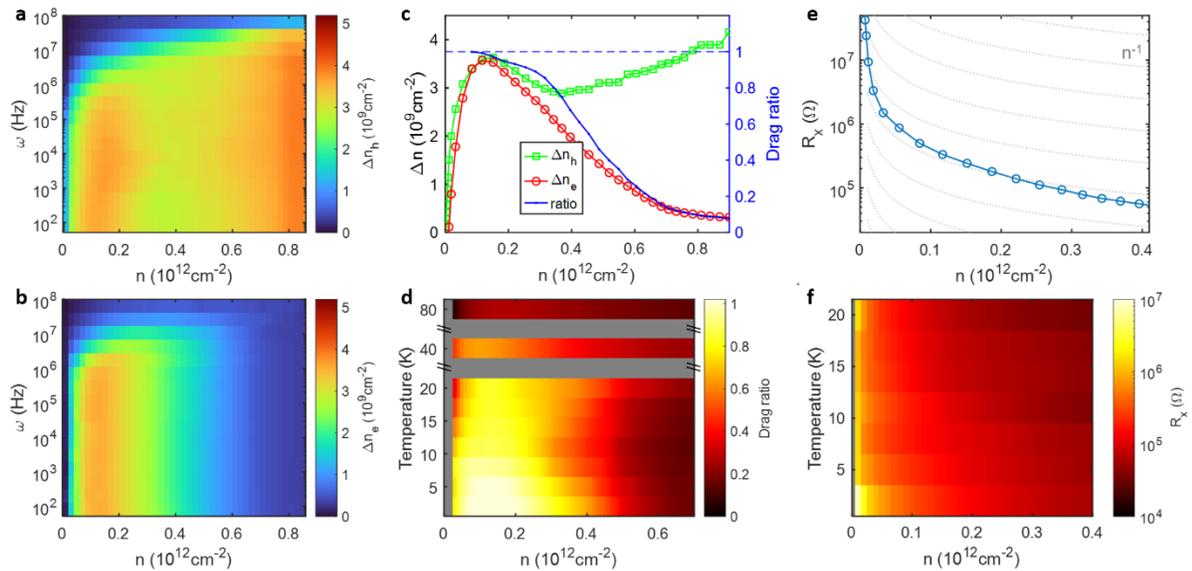

**Fig. 4 | Drag behavior and exciton transport at charge neutrality.**
**a-b**, Density ($n = n_e = n_h$) and frequency ($\omega$) dependence of the drive (hole) and drag (electron) layer density change at net charge neutrality.
**c**, Drive (hole) and drag (electron) layer density change as a function of density $n$. Right axis: the ratio between $\Delta n_e$ and $\Delta n_h$.
**d**, Density and temperature dependence of the drag ratio at net charge neutrality.
**e**, Fitted exciton resistance $R_x$ as a function of $n$.
**f**, Density and temperature dependence of the exciton resistance.

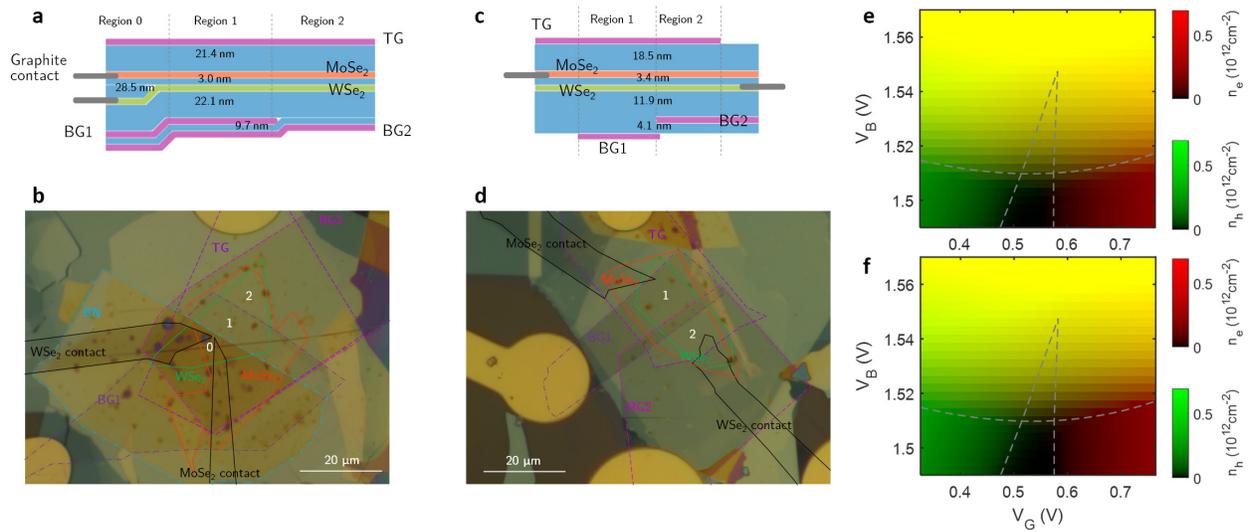

**Extended Data Fig. 1 | Detailed device structure.**
**a-b**, Schematic cross-section and an optical microscope image of device D1.
**c-d**, Schematic cross-section and an optical microscope image of device D2.
**e-f**, Doping phase diagrams of D1 in region 1 (e) and region 2 (f), which are homogeneous down to mV level.

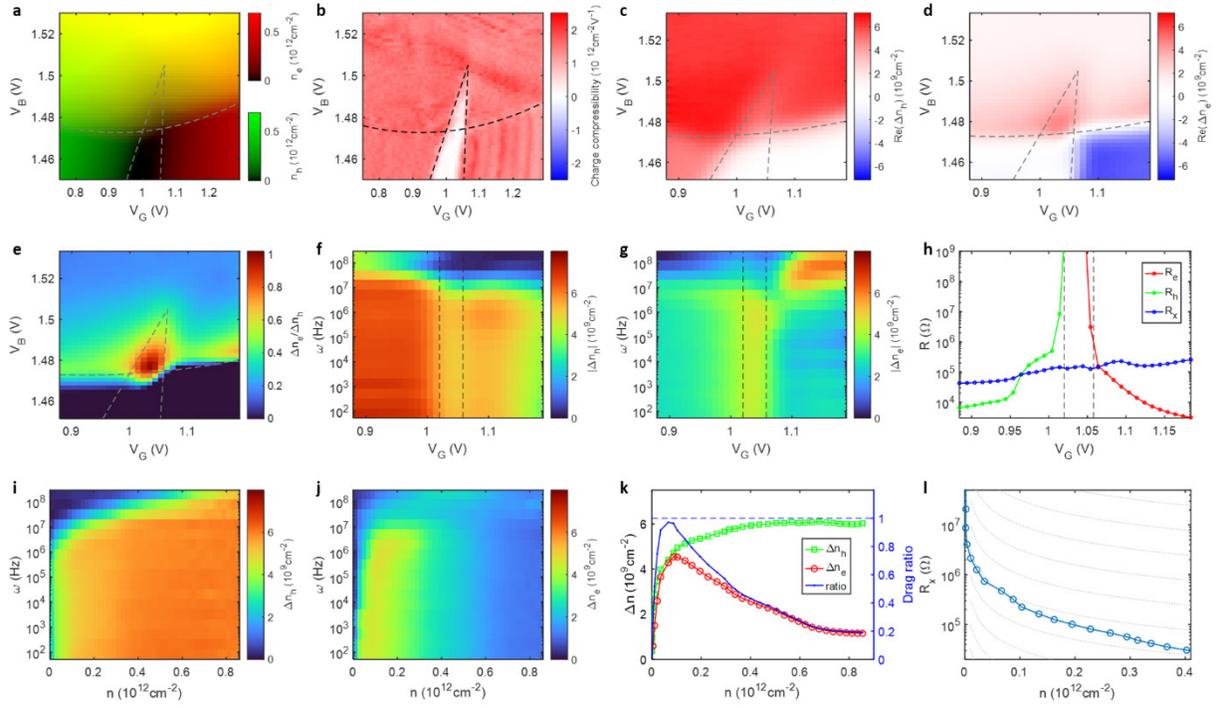

**Extended Data Fig. 2 | Main results from device D2.**
**a**, Carrier doping phase diagram of the electron-hole bilayer.
**b**, Charge compressibility from numerical partial derivative of electron-hole density imbalance with respect to gate voltage.
**c-d**, Density change of the drive layer (c) and the drag layer (d).
**e**, Drag ratio as a function of gate and bias voltages.
**f-g**, Drive (hole) and drag (electron) layer density change under different driving frequencies at constant $V_B = 1.482$V.
**h**, Fitted resistance of free electron ($R_e$), free hole ($R_h$) and interlayer exciton ($R_x$), as a function of $V_G$.
**i-j**, Density ($n = n_e = n_h$) and frequency dependence of the drive (hole) and drag (electron) layer density change at net charge neutrality.
**k,** Drive (hole) and drag (electron) layer density change as a function of $n$. Right axis: the ratio between $\Delta n_e$ and $\Delta n_h$.
**l**, Fitted exciton resistance $R_x$ as a function of $n$.

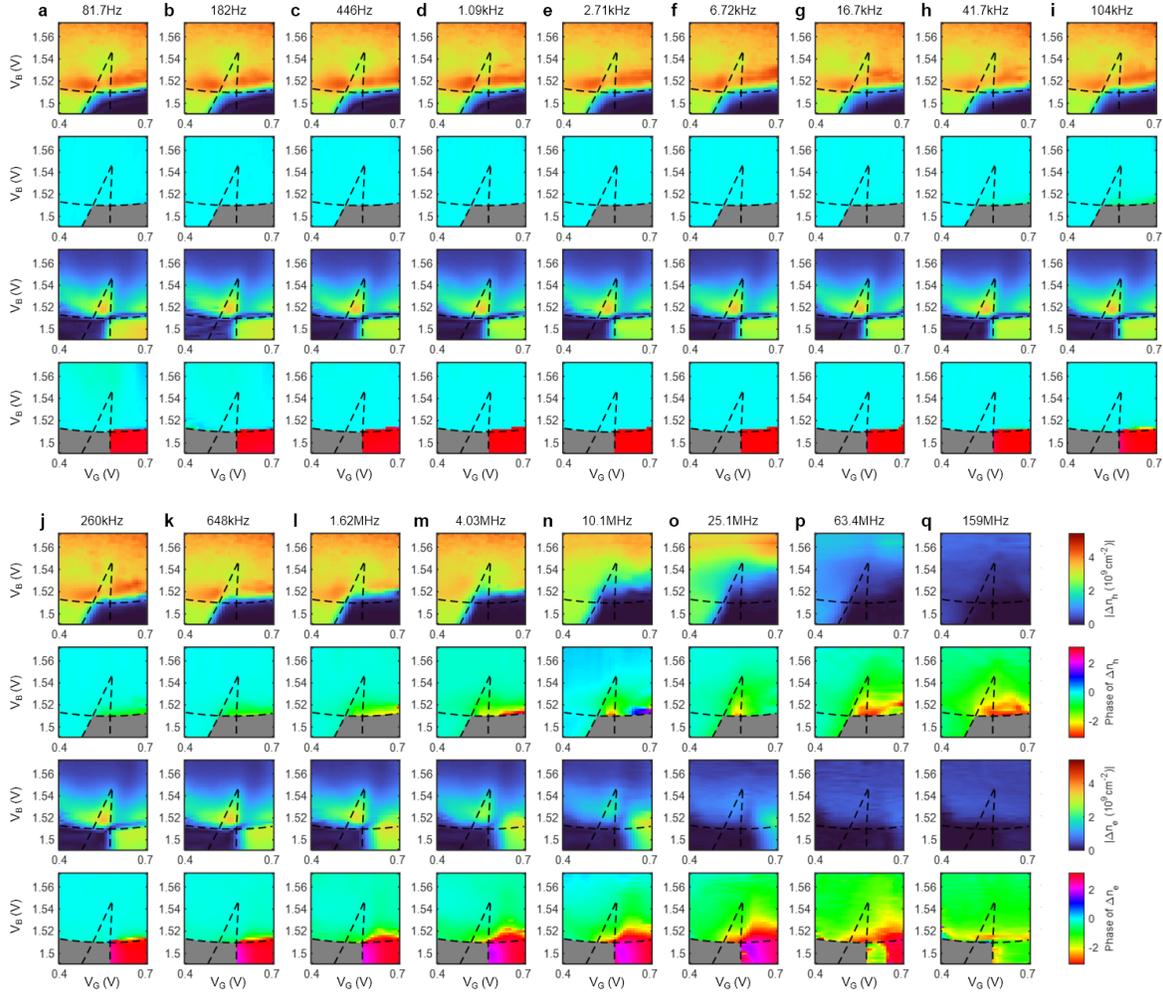

**Extended Data Fig. 3 | Full frequency-dependent dataset.**
**a-q**, Change density change as a function of $V_G$ and $V_B$ at different driving frequencies, showing the amplitude and phase of $\Delta n_h$ and $\Delta n_e$. The top two rows show the hole amplitude and phase and the bottom two rows show the electron amplitude and phase.

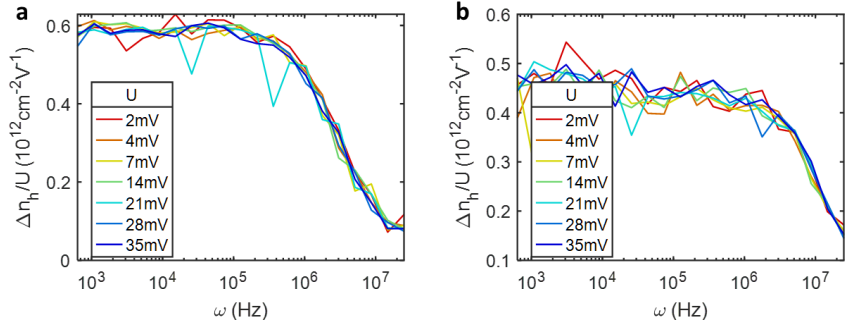

**Extended Data Fig. 4 | Linear exciton transport.**
**a-b**, Modulation amplitude dependence of exciton transport at two typical densities $n_e = n_h = 0.15 \times 10^{12}\,\text{cm}^{-2}$ (a) and $n_e = n_h = 0.3 \times 10^{12}\,\text{cm}^{-2}$ (b). After normalizing for the modulation amplitude, the density change per unit driving voltage decays at the same frequency, indicating the exciton resistance $R_x$ does not depend on exciton current.

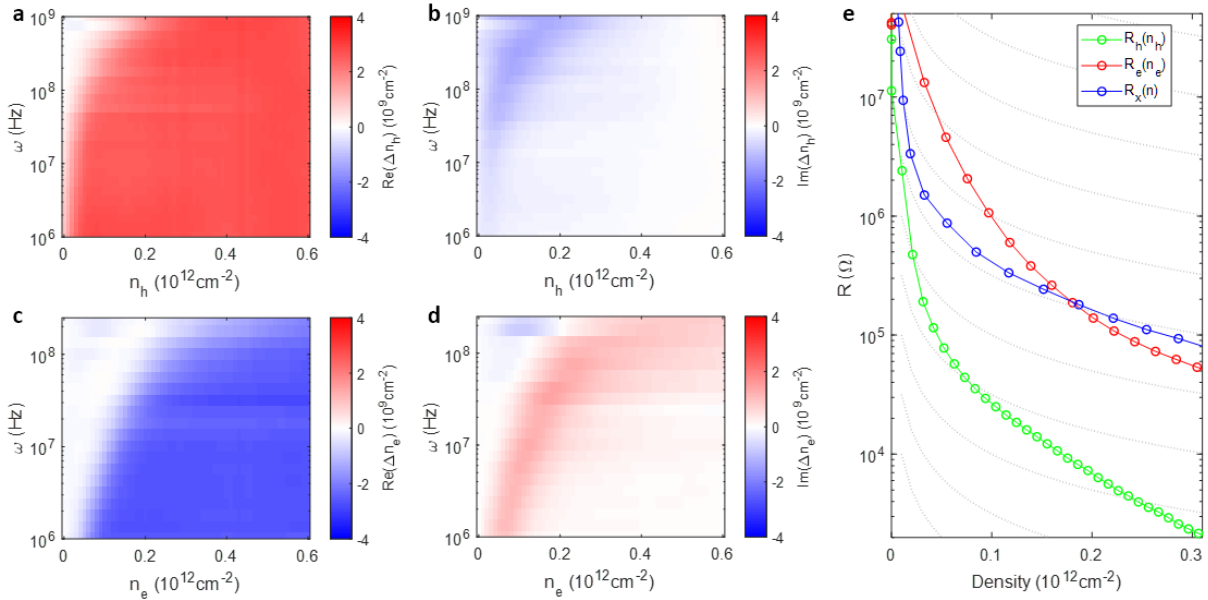

**Extended Data Fig. 5 | Charge transport in the single-charge regime.**
**a-b,** Real and imaginary part of $\Delta n_h$ as a function of $n_h$, when no electron is present.
**c-d,** Real and imaginary part of $\Delta n_e$ as a function of $n_e$, when no hole is present.
**e,** Fitted resistances as a function of carrier densities. Green line, hole layer resistance $R_h$ as a function of $n_h$ when no electron is present. Red line, electron layer resistance $R_e$ as a function of $n_e$ when no hole is present. Blue line, exciton resistance $R_x$ as a function of $n = n_e = n_h$ at net charge neutrality.